\documentstyle[12pt]{article}
\def\reference{\parskip 0pt\par\noindent\hangindent 0.5 truecm}

\begin{document}

\newcommand{\lta}{\mbox{\small\raisebox{-0.6ex}{$\,\stackrel
{\raisebox{-.2ex}{$\textstyle <$}}{\sim}\,$}}}
\newcommand{\gta}{\mbox{\small\raisebox{-0.6ex}{$\,\stackrel
{\raisebox{-.2ex}{$\textstyle >$}}{\sim}\,$}}}   

\title{Probing the pc- and kpc-scale environment of the powerful radio galaxy
Hercules A.}

\author{Nectaria A. B. Gizani, $^{1}$ 
 M. A. Garrett, $^{2}$ \& J. P. Leahy $^{3}$ 
} 

\date{}
\maketitle

{\center 
$^1$ Centro de Astronomia e Astrof\'{i}sica da Universidade de
Lisboa, Observat\'{o}rio Astron\'{o}mico de Lisboa, Tapada da Ajuda, Lisboa,
Portugal, 1349-018\\ngizani@oal.ul.pt\\[3mm]
$^2$ Joint Institute for VLBI in Europe, Postbus 2, Dwingeloo, The
Netherlands, 7990 AA\\garrett@jive.nl\\[3mm]
$^3$ Jodrell Bank Observatory, University of Manchester, Nr
Macclesfield, UK, Cheshire SK 11 9DL\\jpl@jb.man.ac.uk\\[3mm] }

\begin{abstract}

We present the kpc-scale behaviour of the powerful extragalactic radio
source Hercules A and the behaviour of the intracluster gas in which
the radio source is situated. We have found that Hercules A exhibits a
strong Laing-Garrington effect. The X-ray observations have revealed
an extended X-ray emission elongated along the radio galaxy axis. The
estimated temperature of the cluster is kT = 2.45 keV and the central
electron density is $n_{\circ}\,\simeq$ 7.8$\cdot\,10^{-3}$ cm$^{-3}$
which reveals a hot, dense environment in which Hercules A is
situated. From the combined study of the radio and X-ray data we have
estimated a central value of 3 \raisebox{-.6ex}{$\stackrel{\textstyle
<}{\sim}$} $B_{\circ}$( $\mu$G) \raisebox{-.6ex}{$\stackrel{\textstyle
<}{\sim}$} 9.

We also present the most recent results from the analysis of the radio
data on the pc-scale structure of the radio galaxy, observed at 18 cm
by the EVN-MERLIN array. A faint but compact radio source, coincident
with the optical centre of Hercules A was detected by the EVN at 18
mas resolution. The total flux density of the EVN core is 14.6
mJy. Its angular size is 18$\times$7 mas with a position angle of
$\simeq 139^{\circ}$. There is also evidence for extended emission
in the NW-SE direction, most probably from the eastern pc-scale
jet. If this is true then there is a misalignment between the
direction of the pc-eastern and the aligned kpc-scale jets of $\simeq
35^{\circ}$.  

\end{abstract}


\bigskip

\section{Introduction}

\begin{figure}
\centering
\setlength{\unitlength}{1cm}
 
\begin{picture}(6,6)

\put(-5,10){\includegraphics{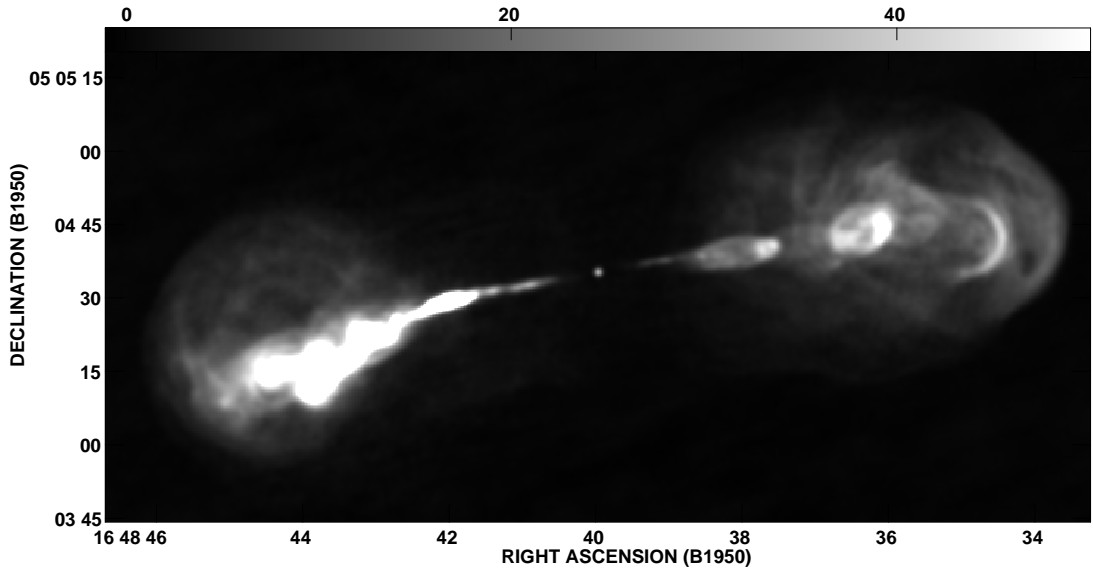}}
\end{picture}
\caption{A grey scale map of Hercules A with Grey scale flux range
-1.0 to 50.0 mJy/beam. The map is at 18 cm, at 1.4 arcsec
resolution. The presence of semi rings is apparent mostly in the
western lobe. The eastern jet exhibits hellical-like features, which
could be ring-like but seen in projection.}
\label{fig1}
\end{figure}

Hercules A is an extended source at a low redshift of z = 0.154 with
total radio luminosity $\sim 3.8\times10^{\footnotesize
37}\normalsize$ W between 10 MHz and 100 GHz. Its power density at 5
GHz is P$_{5\,GHz}$ = 6.9$\times10^{25}$ WHz$^{-1}$sr$^{-1}$ (assuming
H$_{\circ}$=65 kms$^{-1}$ Mpc$^{-1}$ and q$_{\circ}$=0). It is
classified as intermediate between the two FR classes (Dreher \&
Feigelson, 1984).  With a linear size of 540 kpc and no compact
hotspots, Hercules A posesses almost symmetrically extended lobes and
two jets which are quite different in appearance (see Figure
\ref{fig1}). The western jet shows partial or full ring-like features
that form a linear sequence heading from the core to the lobes. The
eastern jet has the highest luminosity found so far in a radio galaxy
associated with an elliptical galaxy (Dreher \& Feigelson, 1984).

Hercules A is identified with a very elongated cD galaxy (e.g. Sadun
et al., 1993, Baum et al., 1996), with absolute magnitude $-23.75$ at
the R-band. The HST observations (Baum et al., 1996) have shown that
the radio source is associated with the diffuse low surface brightness
elliptical galaxy and not with the smaller and brighter companion. The
cluster in which the host galaxy is embedded is of a typical Abell
richness 0 to 1 (Abell 1958).

\section{The kpc-scale environment}

Gizani (1997) has made extensive radio total intensity and polarization
multiband, multiconfiguration observations of Hercules A (Gizani \&
Leahy, 2001a, see also Gizani \& Leahy, 1999) using the VLA, and also
ROSAT X-ray observations. The observations were at 1665, 1435, 1365
and 1295 MHz, in the A-, B-, C-configurations, at 8465 and 8415 MHz in
the B-, C- and D- configurations and we have retrieved and reanalysed
the 4873 MHz data (B-, C-, D-configurations). The purpose of these
multifrequency observations was to map the Faraday rotation field at
high resolution, 1.4 arcsec, combined with the X-ray data on the gas
distribution in order to map the magnetic field of the cluster.

The analysis has shown that Hercules A exhibits a strong
Laing-Garrington effect (Laing, 1988; Garrington et al., 1988), i.e. the
western side of the radio emission is more depolarized than the
eastern side. The projected magnetic field closely follows the edge of
the source, the jets, and the ring structures in the lobes and the
field pattern in the two lobes is broadly similar.
 
Using all 7 frequencies of our radio observations, we have found that
Hercules A is a steep spectrum radio source with spectral index
$\alpha \simeq -1.6$ (we assume that the flux density vary as $S_{\nu}
\propto \nu^{\alpha}$, $\alpha < 0$). Also the more depolarized
western side has a steeper spectrum than the less depolarized eastern
side. The core is optically thin, with $\alpha \simeq -1.20$. The jets
and the ring-like features have spectral index $\simeq -1.1$, while
the surrounding lobes are much steeper, $\simeq -1.6$, which is some
evidence for spectral curvature.

The X-ray emission of the intracluster gas surrounding Hercules A was
observed with the ROSAT PSPC and HRI detectors. These observations
have revealed an extended emission elongated along the radio galaxy
axis. The estimated central electron density of $n_{0} \approx 7.8
\times 10^{-3}$ cm$^{-3}$ reveals a dense environment in which the
radio source is situated. Our spectral fitting with hydrogen column
density of $N_{H} = 6.2 \times 10 ^{20}$ cm$^{-2}$ gives a temperature
estimate of kT = 2.45 keV. The unabsorbed luminosity of the cluster is
$3.16\times 10^{37}$~W. The 0.1--2.4 keV luminosity of the core is
$2.0 \times 10^{36}$~W.

From the combined radio and X-ray analysis, by model fitting the
Faraday dispersion profile from the radio data and the surface
brightness profile from the X-ray data, we have found that the
depolarization is mainly caused by a centrally condensed medium in
which Hercules A is embedded at $\simeq$ 50$^{\circ}$ to the line of
sight (Gizani \& Leahy 1999). Most probably this medium is the `X-ray'
hot gas. According to Unified Schemes the relativistic beaming is
causing the depolarization asymmetry seen between the two sides of the
radio emission: The western weak jet and associated lobe is behind the
bulk of the gas while the bright eastern jet and lobe are in front. As
the radio emission from the lobes travels through this material it is 
affected by Faraday rotation which causes the depolarization. Hence
the further lobe, on the counter-jet side, is more depolarized.

The external magnetic field is decreasing with
radius and Gizani (1997) has estimated a central value of 3
\raisebox{-.6ex}{$\stackrel{\textstyle <}{\sim}$} B$_{\circ}$( $\mu$G)
\raisebox{-.6ex}{$\stackrel{\textstyle <}{\sim}$} 9 (see also Gizani
\& Leahy 1999).

\section{The pc-scale environment}

The VLA data provide us with a good understanding of the kpc-scale
radio structure of Hercules A. There seems to be a symmetry between
the ring-like features of the western jet and the bright helical-like
structures of the eastern jet. The features form a sequence which
suggests successive ejections from the active nucleus.  On the whole
the observed wiggles in the large scale jets and the overall symmetry
of the envelope of the radio structure may suggest precession or
wobbling of the radio axis. The flux of the core is 41 mJy at 18 cm at
1.4 arcsec resolution. The inner jets of Hercules A together with its
optically thin, steep spectrum core, were unresolved with the VLA.

Taking the above into account we have tried to study the pc-scale
environment of this powerful radio source using the EVN+MERLIN
interferometer at 18 cm. The main goal of the EVN and joint MERLIN
observations (Gizani, Garrett \& Leahy, 2000, and 2001) was to try and
detect any pc-scale jet emission on both sides of the central engine,
to map the transition region between them and the kpc-scale jets with
the highest possible resolution and to resolve the inner jets in
detail. We have observed Hercules A for 11 hours employing phase
referencing since the emission from the area of the core was so weak.

\begin{figure}
\centering
\setlength{\unitlength}{1cm}
 
\begin{picture}(6,6)

\put(-4.5,-6){\includegraphics{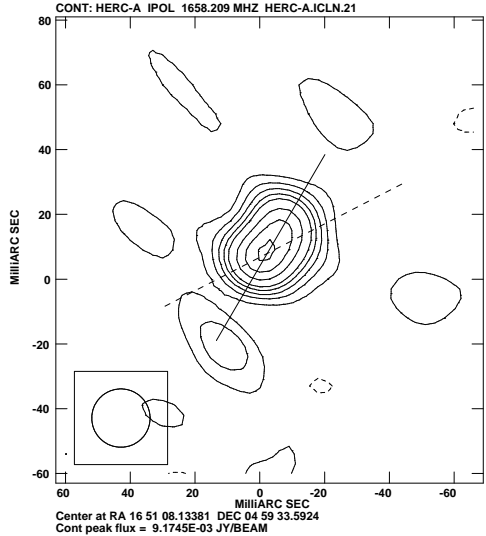}}
\end{picture}
\caption{The EVN contour map of the core region of Hercules A. The
dashed line indicates the direction of the kpc-scale jets. The solid
line shows the direction of the possible NW/SE emission with the
current observations (see text). Contours are at levels $-$1, 1, up to
10 with increaments of 1, starting at 2.5$\times 10^{-5}$ Jy/beam.}
\label{fig2}
\end{figure}

There were problems with the MERLIN observations, due to the diffuse,
extended structure of this bright source. We have therefore combined
our VLA data with the MERLIN data at 18 cm, in order to obtain
information of the inner region of the radio source. This work is in
progress. Figure~\protect{\ref{fig2}} shows the EVN detection of the
core region.

A compact core is detected with the EVN at 18 mas resolution. Using a
model with one gaussian component we have found the size of the core
to be $18 \times 7$ mas at position angle $139^{\circ}$. The error
estimate based on rms is 3.6$\times 10^{-4}$ Jy/beam. The flux density
is $\simeq$ 15 mJy, which implies a radio power of 3.6 $\times
10^{22}$ W\,Hz$^{-1}$sr$^{-1}$.  The implied brightness temperature of
the compact core is $\simeq 2 \times 10^7$ K. The core is still not
resolved out at 18 mas and very weak. There is an indication of a real
emission from the core region in the NW/SE direction (see
Figure~\protect{\ref{fig2}}). If this is true then it could be
interpreted as the emission from the core and one-sided jet: According
to Doppler beaming the mas scale jet is always on the same side as the
brighter kpc-scale jet (eastern jet in our case). Therefore the core
should be located at NW of the observed emission while the SE emission
is from the eastern pc-scale jet. If this is indeed the case, then
there is a misalignment between the direction of the eastern pc-scale
jet, at least, with the larger scale jets of $\sim 35^{\circ}$.  This
misalignment angle which is relatively high compared to the ones found
in other powerful radio galaxies (Cyg A, Krichbaum et al., 1998,
3C\,109, Giovannini et al., 1994 for example), is characteristic of
quasars in the CSS class with steep spectrum cores (Fejes et al., 1992
for example for the OVV quasars 3C\,216 and 3C\,446 with 'superluminal
behaviour', and references therein). This result could be consistent
with the suggestion that powerful radio galaxies are the unbeamed
counterparts of quasars (Saripalli et al., 1997).

\section{Conclusions and Future Work}

In kpc-scales we have found that Hercules A exhibits a strong
Laing-Garrington effect. The depolarization asymmetry between the two
sides can be explained by Hercules A being embedded in a centrally
condensed medium at a substantial angle to the line of sight. This
medium should be the X-ray emitting gas. The more depolarized western
side has a steeper spectrum than the less depolarized eastern side;
also the jets and rings in the western lobe have a much less steep
spectrum than the lobe material, suggesting that we are witnessing a
renewed outburst from the active nucleus. Our X-ray observations of
the intracluster gas in the Hercules A cluster have revealed an
extended X-ray emission and a possible weak nuclear component. The
cluster exhibits a significant concentration of gas towards the
centre; It is hot and dense. We estimated the central value of the
external magnetic field strength by model fitting the radio and X-ray
data. We found that the magnetic field is decreasing with radius and
its estimated central value is 3
\raisebox{-.6ex}{$\stackrel{\textstyle <}{\sim}$} $B_{\circ}$( $\mu$G)
\raisebox{-.6ex}{$\stackrel{\textstyle <}{\sim}$} 9.

In pc-scales the results came from the EVN detection at 18 cm. About
30\% of the VLA flux from the core area at 18 mas is detected with
angular size 18$\times$7 mas, position angle $\simeq 139^{\circ}$ and
error estimate based on rms 3.6$\times 10^{-4}$ Jy/beam. Its
brightness temperature is 2$\times 10^{7}$ K.  The core is still not
resolved out and very weak.  There is evidence for emission extended
in the SE direction. This emission is probably from the eastern
pc-scale jet, due to Doppler boosting. Then one can conclude that
there is a misalignment between the pc- and kpc-scale jet (at least
the eastern one) of $35^{\circ}$. Not much is known about the
curvature of the pc-, kpc- scale jets for steep spectrum sources
versus redhift. This study is within our scientific interest.

For the future, since there is a possibility the counter jet is being
obscured by a forground absorption and a possible toroidal or
circum-nuclear disk, we are planning optical and infrared observations
as well as HI absorption measurements. The latter could prove the
presence of a foreground absorption with the galactic column density
assumed in our analysis. UV observations, to use together with other
wavelength observations, as diagnostic of the ISM around the central
region together with the study the UV emission from nuclear region,
are also within our plans. High sensitivity VLBI observations at lower
frequencies could determine the extent of the eastern pc-scale jet and
could search for the presence of a western counterjet. Simultaneous
multifrequency VLBI observations could be the method of probing the
parsec-scale environment of Her A. 

It could also be interesting to see whether the VLBI core of Her A is
variable, like Cygnus A (Krichbaum et al., 1998) and 3C\,338 (Feretti
et al., 1993) for example.

\section*{Acknowledgments}

NG is grateful to the LOC of this very interesting workshop for giving
her {\underline most} of the financial support without which, her
participation to the workshop would have been impossible.  NG
acknowledges the European Commission's TMR "Access to Large-Scale
Facilities" programme, under contract ERBGCHT950012 in order to reduce
the EVN+MERLIN data on Hercules A. NG would also like to acnowledge
her postdoc grant PRAXIS XXI/BPD/18860/98 from the Funda\c c\~ao para
a Ci\^encia e a Tecnologia, Portugal, which covered the rest of the
expenses her participation to this workshop and supports her work. The
European VLBI Network is a joint facility of European and Chinese
radio astronomy institutes funded by their national research councils.

\section*{References}

\reference 
Abell, G.O., ApJSupp, 1958, 3, 211
\reference 
Baum, S. A., O'Dea, C. P., De Koff, S., Sparks, W., Hayes, J. J. E., Livio, M. \& Golombek, D. ApJL., 1996, 465, 5 
\reference 
Dreher, J. W. \& Feigelson, E. D. Nat., 1984, 308, 43 
\reference
Fejes, I., Porcas, R. W. \& Akujor, Chidi. E., 1992, A.A., 257, 459
\reference 
Feretti, L., Comoretto, G., Giovannini, G., Venturi, T. \& 
Wehrle, A.E., ApJ, 1993, 408, 446 
\reference 
Garrington, S. T., Leahy, J. P., Conway, R. G. \& Laing, R. A., Nat., 1988, 331, 147 
\reference 
Giovannini, G., Feretti, L., Venturi, T., Lara, L.,
Marcaide, J., Rioja, M., Spangler, S. R. \& Wehrle, A. E., 1994, ApJ, 435, 116
\reference
Gizani, N. A. B. 1997, 'Environments of Double Radio Sources 
associated with Active Galactic Nuclei',  PhD thesis, Jodrell Bank 
Observatories, University of Manchester, UK. 
\reference 
Gizani, N. A. B. \& Leahy, J. P. 1999, In 
{\it EVN/JIVE.\ Symp.\ No. 4\/} (ed.\ Garrett, M. A., Campbell, R. M. 
\& L. I. Gurvits). NewAR, Elsevier Science Ltd publishers, vol. 43, p. 639   
\reference 
Gizani, N.A.B., Garrett, M. \& Leahy, J.P.,
2000, ``The Behaviour of the central engine of the powerful
extragalactic radio source Hercules A'', in 'Proceedings of the 
5th European VLBI Network Symposium', J.E. Conway, A.G. Polatidis, 
R.S. Booth and Y.M. Pihlstr\"{o}m, (eds.), 
Onsala Space Observatory Publications ISBN 91-631-0548-9, p. 19 
\reference 
Gizani, Nectaria A. B. \& Leahy, J. P. MNRAS, 2001a, in prep. 
\reference 
Gizani, Nectaria A. B. \& Leahy, J. P. MNRAS, 2001b, in prep. 
\reference 
Gizani, Nectaria A. B., Garrett, M. A. \& Leahy, J. P., 
MNRAS, 2001, in prep.   
\reference 
Krichbaum, T.P. et al., A.A., 1998, 329, 873 
\reference 
Laing, R. A.  Nat., 1988, 331, 149   
\reference 
Sadun, A.C. \& Hayes, J.J.E., PASP, 1993, 105, 379 
\reference
Saripalli, L. et al., 1997, A.A., 328, 78
\end{document}